\pdfoutput=1
\documentclass[
    reprint,
    aps,
    pre,
    amsmath,
    amssymb,
    superscriptaddress,
    nofootinbib
]{revtex4-2}

\usepackage{newtxtext}

\usepackage{graphicx}
\usepackage[caption=false]{subfig}
\usepackage{placeins}

\usepackage[normalem]{ulem}
\usepackage{color}
\usepackage{lineno}

\usepackage[
    colorlinks=true,
    citecolor=blue,
    linkcolor=blue,
    urlcolor=blue
]{hyperref}

\newtheorem{theorem}{Theorem}

\begin{document}

\title{Percolation on interdependent one-dimensional long-range networks}

\author{Adar Sharabi}
\affiliation{
Department of Natural Sciences,
The Open University of Israel,
Ra'anana, Israel
}

\author{Amir Bashan}
\affiliation{
Department of Physics,
Bar-Ilan University,
Ramat Gan, Israel
}

\author{Sergey V. Buldyrev}
\affiliation{
Department of Physics,
Yeshiva University,
New York, New York, USA
}

\author{Guy Amit}
\email{guyam@openu.ac.il}
\affiliation{
Department of Natural Sciences,
The Open University of Israel,
Ra'anana, Israel
}

\date{\today}

\begin{abstract}
We study the emergence of a mutual giant component in a multiplex network comprised of two one-dimensional long-range networks on a lattice. The probability that two nodes on one lattice, with a distance $r$ between them, are connected with an edge falls algebraically with distance as $Cr^{-(1+\sigma)}$, where $0<\sigma<1$ and $0<C\leq1$. A mutual giant component exists if $C>C_c^{\mathrm{mul}}$, where $C_c^{\mathrm{mul}}(\sigma)$ is a critical value of $C$ that is dependent on $\sigma$. We find rigorous lower bounds on $C_c^{\mathrm{mul}}$, and a tree approximation whose estimate of $C_c^{\mathrm{mul}}$ agrees with simulations within $0.6\%$ for $\sigma\leq0.5$. The multiplex network is significantly less stable than the single network case, as is evident by the larger value of $C_c^{\mathrm{mul}}$, which is up to $2.4$ times the single-layer value. Our simulations indicate that the order of the transition depends on $\sigma$. For $\sigma\leq0.35$   the mutual giant component collapses abruptly, in a first-order phase transition, while for $\sigma\geq0.4$ the transition is continuous. When the dependency links connect random pairs of nodes, the transition is first order for all $\sigma$.
\end{abstract}

\maketitle

\section{Introduction}
\label{sec:intro}

In this work, we ask the following question: If two networks depend on each other, and the links within each network are long-ranged, under what conditions will the coupled system have a giant component? And when this condition is violated, does the system collapse abruptly or gradually?

Networks are a common framework for the description of complex systems, where the components of the system are the nodes, and the interactions between them are the links~\cite{albert2002,newman2010book,cohen2010book}. A giant component (GC) is a connected cluster of nodes whose size remains greater than $cN$, where $0<c\leq1$ is some constant, for any $N\to\infty$. The existence of a GC signifies the ability of the network to function properly~\cite{molloy1998,strogatz2001,newman2001,newman2002}, and its emergence is a percolation phase transition~\cite{stauffer1994,cohen2010book}.

In many real-world systems, however, a network does not function in isolation, but depends on other networks~\cite{rinaldi2001,vespignani2010}. For example, the stations of a power grid depend on the communication network that controls them, while the communication network depends on the power grid for its electricity~\cite{rinaldi2001,buldyrev2010}. Similar dependencies exist between transportation networks~\cite{cardillo2013}, between different social relations of the same individuals~\cite{szell2010}, between financial institutions~\cite{huang2013}, and between the structural and functional networks of the brain~\cite{dedomenico2017}. Such systems are modeled as interdependent, or multilayer, networks~\cite{kivela2014,boccaletti2014,dedomenico2013,bianconi2018}. In the model of Buldyrev \textit{et al.}~\cite{buldyrev2010}, two networks are coupled by \textit{dependency links}, such that a node can function only if the node it depends on in the other network functions as well. A failure of a node therefore propagates back and forth between the networks, in a cascade of failures. As a result, interdependent networks are significantly more vulnerable than single networks, and their mutual giant component may collapse abruptly, in a first-order phase transition~\cite{buldyrev2010,parshani2010,gao2011,gao2012,baxter2012,son2012,radicchi2015}. The transition is not always first order, however. It becomes continuous when only a small enough fraction of the nodes have dependency links~\cite{parshani2010}, and also when the networks are embedded in space and the dependency links are short, as discussed below.  An important special case is a \textit{multiplex} network, in which the layers share the same set of nodes, and each node depends on its own copy in the other layer~\cite{baxter2012,boccaletti2014,danziger2019}.

Many of these networks are also embedded in space~\cite{barthelemy2011,gross2022book}, and spatial embedding changes the nature of the transition. When the links within each network are short-ranged, a multiplex of two spatial networks, where each node depends on the node at the same position, undergoes a continuous transition. This is the case for two lattices~\cite{son2011,li2012,grassberger2015}, but the networks do not have to be lattices: the transition is also continuous for spatial networks whose links have a short characteristic length, and it becomes discontinuous when the characteristic length of the links is increased~\cite{danziger2016}.   When the dependency links connect nodes that are far apart, or randomly chosen nodes, the transition becomes first order, and the system is extremely vulnerable~\cite{li2012,bashan2013}  . In spatial interdependent networks, even a localized failure can spread through the system and collapse it~\cite{berezin2015,danziger2016,vaknin2017}, and the mixed-order transitions of such systems exhibit fractal fluctuations~\cite{gross2022}. These effects are not only theoretical, as interdependent networks were recently realized in the laboratory, using networks of disordered superconductors~\cite{bonamassa2023}.

Real-world spatial networks, however, are not restricted to links between nearby nodes. In many cases, the probability of a link decays as a power law of the distance, as in the network of global airlines~\cite{daqing2011}. Such long-range (LR) interactions appear in many areas of physics, including gravitational systems, plasma physics, hydrodynamics, atomic physics, and nuclear physics~\cite{campa2009,campa2014,defenu2023}, and they were also used to model disease spreading~\cite{janssen1999,linder2008}, social aggregation~\cite{solomon2000}, and navigation and transport in networks~\cite{kleinberg2000,li2010}. Long links have a strong effect even when they are placed at random, regardless of the distance: as shown by Watts and Strogatz~\cite{watts1998}, a small fraction of such shortcuts in an ordered network changes its collective behavior dramatically.

The simplest model of a network with LR links is percolation on a one-dimensional lattice with LR interactions~\cite{schulman1983,newman1986,aizenman1986,gori2017}, where each pair of nodes at a distance $r$ is connected with probability $p_r=C/r^{1+\sigma}$. As in the Ising model with LR interactions~\cite{dyson1969,kac1969,fisher1972}, the LR links allow a phase transition in one dimension, which is impossible with short-range links alone. Despite its simplicity, the critical value of this model is not known exactly. The available results are bounds~\cite{schulman1983} and numerical estimates~\cite{gori2017}, and its critical behavior was established rigorously only recently~\cite{berger2002,hutchcroft2024,hutchcroft2025II}. In a previous work, we showed that this model also describes the percolation of fractal spatial networks with LR interactions, whose nodes are positioned in space by a L\'evy flight~\cite{amit2023}.

Here, we combine the two conditions, and study a multiplex network of two one-dimensional LR networks, where node $i$ of one layer depends on node $i$ of the other layer~\cite{son2024poster}. This model is the LR analog of the multiplex of lattices~\cite{son2011,li2012}. The parameter $\sigma$ allows us to tune the structure of the network continuously, from almost random for small $\sigma$, where most links are long, to almost one-dimensional as $\sigma\to1$. We find rigorous lower bounds on the critical value $C_c^\mathrm{mul}$, and a tree approximation that predicts it within $0.6\%$ for $\sigma\leq0.5$. Our simulations indicate that the transition is first order for $\sigma\leq0.35$, and continuous for $\sigma\geq0.4$. The tree approximation, in contrast, predicts a first-order transition for all $\sigma$, so that for $\sigma\geq0.4$ it gives the critical value but not the order of the transition.

The paper is organized as follows. In Sec.~\ref{sec:model} we define the model. In Sec.~\ref{sec:bounds} we present lower bounds on $C_c^\mathrm{mul}$, and in Sec.~\ref{sec:tree} we derive the tree approximation. The simulations are presented in Sec.~\ref{sec:sim}, and discussed in Sec.~\ref{sec:discussion}. The proofs of the bounds are given in Appendix~\ref{app:bounds}.

\section{Model}
\label{sec:model}
\begin{figure}[t]
    \centering
    \includegraphics[width=\columnwidth]{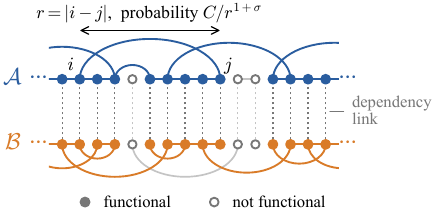}
    \caption{Schematic description of the model. Two layers, $\mathcal{A}$ and $\mathcal{B}$, are infinite one-dimensional lattices. In each layer, two nodes $i$ and $j$ at a distance $r=|i-j|$ are connected with probability $p_r=C/r^{1+\sigma}$, independently of all other pairs and of the other layer. Node $i$ of $\mathcal{A}$ is connected with a dependency link to node $i$ of $\mathcal{B}$ (dashed lines), such that node $i$ of $\mathcal{A}$ is functional only if its corresponding node in $\mathcal{B}$ is also functional. Filled circles: an example of a set of functional nodes, which form a connected cluster in each of the two layers. Empty circles: nodes that are not functional, since in at least one of the layers they are not connected to this cluster. The links of these nodes (light gray) are removed.}
    \label{fig:model}
\end{figure}

The multiplex network is constructed from two layers, $\mathcal{A}$ and $\mathcal{B}$, that are independent copies of random graphs on $\mathbb{Z}$ (see Fig.~\ref{fig:model}). A pair of nodes in the same layer, $i$ and $j$, ($i\neq j$) at a distance $r=\left|i-j\right|$ from each other have an edge with probability
\begin{equation}
    p_r=\frac{C}{r^{1+\sigma}},
    \label{eq:pr}
\end{equation}
with $0<\sigma<1$ and $0<C\leq 1$. In addition, each node, $i$, in layer $\mathcal{A}$ is connected with a \textit{dependency link} to its corresponding node in layer $\mathcal{B}$. We denote the largest connected component of a layer by LCC.

A node, $i$, is \textit{functional} if it satisfies the following two conditions
\begin{enumerate}
    \item It is a member of the LCC of its layer, formed by functional nodes only.
    \item It is connected with a dependency link to a functional node.
\end{enumerate}
The definition is recursive. In practice, the set of functional nodes, $M$, is found by the cascade of failures as in Ref.~\cite{buldyrev2010}: every node that is not in the LCC of its layer fails together with its partner in the other layer, and this is repeated until no further node fails (Sec.~\ref{sec:sim}).
The number of nodes in each layer is $N$, and we consider the case of $N\to\infty$. A mutual giant component (MGC) exists if both layers have an infinite connected cluster of functional nodes.

The case of a single layer has been extensively studied \cite{stephen1981,zhang1983,schulman1983,newman1986,aizenman1986,imbrie1988,berger2002,gori2017,duminil2024,hutchcroft2025II}. For $0<\sigma<1$, an infinite connected cluster exists if and only if
\begin{equation}
    C>C_c^\mathrm{single}\geq C_\mathrm{S}=\frac{1}{2\zeta(\sigma +1)},
    \label{eq:CS}
\end{equation}
where $\zeta(s)$ is the Riemann zeta function,
\begin{equation}
    \zeta(s)=\sum_{n=1}^\infty \frac{1}{n^s}.
\end{equation}
The lower bound $C_\mathrm{S}$ was found by Schulman~\cite{schulman1983}. Newman and Schulman also showed that $C_c^\mathrm{single}<1$~\cite{newman1986}. For $\sigma\geq1$, a GC exists only if $C=1$, i.e., if all nearest neighbors are connected~\cite{newman1986,aizenman1986}.
Gori \textit{et al.}~\cite{gori2017} found precise numerical estimates of $C_c^\mathrm{single}$ from the crossing points of the rescaled mean cluster size, $SN^{-\sigma}$, for different system sizes, up to $N=5.12\times10^5$ (see Table~\ref{tab:sim}).

The order parameter of the system is the relative size of the largest connected component, $P_\infty$. In the single layer case, for a fixed $\sigma$, as $C$ is increased from $C<C_c^\mathrm{single}$ to $C>C_c^\mathrm{single}$ the system undergoes a continuous (second-order) phase transition: there is no infinite cluster at $C=C_c^\mathrm{single}$~\cite{berger2002}, so that the order parameter increases continuously from $P_\infty=0$ to $P_\infty>0$. Near the threshold, $P_\infty\sim(C-C_c^\mathrm{single})^\beta$. For $\sigma<1/3$ the critical behavior is mean-field~\cite{hutchcroft2024}, so that $\beta=1$, and the first derivative of $P_\infty$ jumps from zero to a finite value at $C_c^\mathrm{single}$. For $1/3<\sigma<1$ the value of $\beta$ is not known exactly. The results of Refs.~\cite{hutchcroft2024,hutchcroft2025II} were derived for a closely related model, in which two nodes are connected with probability $1-\exp(-\lambda r^{-(1+\sigma)})$, which has the same behavior at large distances. 

Multiplex networks are less stable and more vulnerable to random attacks than single networks, due to the cascade of failures~\cite{buldyrev2010,parshani2010,baxter2012}. In the multiplex, the critical value cannot be smaller than in the single network case, $C_c^\mathrm{mul}\geq C_c^\mathrm{single}$, since the functional nodes form a connected cluster in each layer separately. Furthermore, since the failure of a node in one layer removes its partner in the other layer, which may disconnect further nodes, the damage is amplified back and forth between the layers. As a result, as the parameter $C$ increases, the system can undergo a \textit{first-order} phase transition, with an abrupt jump in the value of the order parameter, the relative size of the mutual giant component, $P_\infty^\mathrm{mul}$, at $C=C_c^\mathrm{mul}$~\cite{buldyrev2010}. In interdependent random networks the transition is first order~\cite{buldyrev2010}, while in a multiplex of two-dimensional lattices it is continuous~\cite{son2011,li2012}. Our simulations below indicate that in the present model both cases occur, depending on $\sigma$.

\section{Results}
\subsection{Analytical bounds}
\label{sec:bounds}
Even in the single layer case, an exact closed-form formula for $C_c^\mathrm{single}(\sigma)$ is unknown. The best-known analytical result is the lower bound $C_\mathrm{S}$ of Eq.~\eqref{eq:CS}. Similarly, in the multiplex case, we find rigorous lower bounds on the value of $C_c^\mathrm{mul}$. The full derivation is presented in Appendix~\ref{app:bounds}.

We define the following quantities,
\begin{align}
    &\langle k\rangle =2\sum_{r= 1}^\infty p_r=2C\zeta (1+\sigma),\label{eq:k}\\
    &P_+=\prod_{r=1}^\infty (1-p_r),\label{eq:Pplus}\\
    &q=1-P_+^2,\label{eq:q}
\end{align}
where $\langle k\rangle$ is the mean degree of a node in a single layer, $P_+$ is the probability that a node has \textit{no} links to its right, and $q$ is the probability that a node has at least one link in a layer. Note that $q\geq 1-(1-C)^2>C\geq p_r$.

Next, we define
\begin{equation}
    \eta = 1-P_+\prod_{r= 1}^\infty \left(1-\frac{p_r}{q}\right),
    \label{eq:eta}
\end{equation}
which satisfies $q<\eta<1$. The meaning of $\eta$ is the following. A mutual giant component requires a long path of linked nodes in one layer, such that every node of the path also has at least one link in the other layer. For a single node, the probability to have at least one link in the other layer is $q$. For the nodes of a path these events are not independent, since two nodes of the path may be linked to each other in the other layer, and $\eta$ is an upper bound on the probability that the next node of the path has such a link, given that all the previous nodes have one (Appendix~\ref{app:bounds}). Each step of the path therefore costs a factor $\langle k\rangle$ for the link within the layer, and a factor $\eta$ for the link in the other layer.
Then, the following theorems hold,

\begin{theorem}
\label{thm:bound}
If $\langle k\rangle\eta<1$, then there is no mutual giant component. The product $\langle k\rangle\eta$ increases continuously with $C$, from $0$ at $C\to0$ to values above $1$ at $C\to1$. The condition $\langle k\rangle\eta=1$ therefore defines a unique value of $C$, $C_*(\sigma)$, and
\begin{equation}
    C_c^\mathrm{mul}(\sigma)\geq C_*(\sigma)>C_\mathrm{S}=\frac{1}{2\zeta(1+\sigma)}.
\end{equation}
\end{theorem}
The strict inequality holds because at $C = C_S$ we have $\langle k\rangle=1$ and $\eta<1$. Based on this, we find a slightly smaller lower-bound as a closed form expression for the non-existence of a mutual giant component (Appendix~\ref{app:bounds}),
\begin{theorem}
\label{thm:closed}
    There is no mutual giant component if
    \begin{equation}
        2Cz\left[1-\left((1-C)\left(1-\frac{C}{1-e^{-2Cz}}\right)\right)^z\right]<1,
        \label{eq:closed}
    \end{equation}
    with $z=\zeta(1+\sigma)$.
\end{theorem}
The critical value of $C$ based on Theorem~\ref{thm:closed}, $C_{**}$, is within $4\%$ of the value of $C_*$ (see Table~\ref{tab:sim}).

\subsection{Tree approximation}
\label{sec:tree}
The Schulman lower bound for the one-layer network is based on comparison with a tree network (Bethe lattice~\cite{bethe1935,fisher1961}). In a tree network, starting with node $0$, we treat every new node linked to $0$ as belonging to a new, independent, infinite one-dimensional lattice, and connect to it the new nodes with the same probability, $p_r$. For completeness, we repeat the argument here.

Let $\rho$ be the probability that a node belongs to the giant component (so that $\rho$ is the relative size of the giant component of the one-layer network). Look at node $0$ and at one other node, $j$, at a distance $r$. We call the link $0$--$j$ \textit{successful} if it connects node $0$ to the giant component. This requires two things:
\begin{enumerate}
    \item The link exists, which has probability $p_r$.
    \item Node $j$ is itself connected to the giant component by a route that does not come back through node $0$.
\end{enumerate}
Node $0$ belongs to the giant component exactly when at least one of its links is successful.

Next, we approximate the network as a tree. We give the second event a probability $\rho$, and we treat the links to different nodes, $j$, as independent. This would be exact if the network were a tree in which every newly reached node has its own complete, fresh set of possible links. It ignores loops, for instance the fact that two neighbors of node $0$ may be linked to each other.

With this approximation, the link to node $j$ is successful with probability $p_r\rho$ and unsuccessful with probability $1-p_r\rho$. Node $0$ is outside the giant component only if \textit{all} its links are unsuccessful:
\begin{equation}
    1-\rho = \prod_{j\neq0} (1-p_{|j|}\rho)=\prod_{r=1}^\infty\left(1-\frac{C\rho}{r^{1+\sigma}}\right)^2.
    \label{eq:tree1}
\end{equation}
Define $u=C\rho$, and analogous to Eqs.~\eqref{eq:Pplus} and~\eqref{eq:q} we define
\begin{equation}
    P_+(u)=\prod_{r=1}^\infty \left(1-\frac{u}{r^{1+\sigma}}\right),\quad q(u)=1-P_+(u)^2,
    \label{eq:Pu}
\end{equation}
as the probability that \textit{no} link to the right of $0$ is successful, and that \textit{at least} one of the links is successful, respectively. Equation \eqref{eq:tree1} is therefore simply,
\begin{equation}
    \rho=q(u)=1-P_+(u)^2.
    \label{eq:rhoq}
\end{equation}
$\rho=0$ is a trivial solution to Eq.~\eqref{eq:rhoq}. To find other solutions, we first note that $u$ is strictly between $0$ and $1$, since $C\leq1$ and $\rho<1$. Every $u$ between these limits gives one more, nontrivial, solution,
\begin{equation}
    \rho = q(u), \quad C=\frac{u}{q(u)}.
    \label{eq:Cu}
\end{equation}
Reading this the other way round: a non-zero solution exists at a given $C$ only if $C=u/q(u)$ for some $u$. The minimal value of $C$ that fulfills this condition is the critical value of the tree approximation. The function $u/q(u)$ is a monotonically increasing function which is minimal at $u\to0$. This follows from the shape of $q(u)$: every factor $(1-u/r^{1+\sigma})^2$ of $1-q(u)=P_+(u)^2$ is positive, decreasing and convex in $u$, so their product is convex, and $q(u)$ is concave with $q(0)=0$. For such a function the ratio $q(u)/u$, the slope of the chord from the origin, can only decrease with $u$.  In this limit, $P_+(u)^2\approx1-2u\sum_r r^{-(1+\sigma)}=1-2\zeta(1+\sigma)u$ and $q(u)\approx2\zeta(1+\sigma)u$, so that
the non-trivial value of $C$ is $C=C_\mathrm{S}=1/2\zeta(1+\sigma)$ [see the one-layer curve in Fig.~\ref{fig:tree}(b)]. In the case of a single layer, a giant component does not exist if the corresponding tree network also has no giant component, because loops can only reduce the number of nodes that are reached from node $0$. Therefore, $C_\mathrm{S}$ is both the tree approximation of the critical value, and also the lower bound, for the single-layer network.

In the multiplex case, a node is functional only if it is connected to the giant component in both layers, through links to functional nodes (Sec.~\ref{sec:model}). As in the single-layer case, Eq.~\eqref{eq:rhoq}, this leads to a self-consistent equation for $P_\infty$, Eq.~\eqref{eq:Pmul_tree} below, but now with two conditions, one for each layer.  The conditions for a link to be \textit{successful} are now the following:

\textbf{Condition (A).} Take a node $j$ at distance $r$. The $\mathcal{A}$-link $0$--$j$ is now called successful if
it exists (probability $p_r$) and node $j$ is functional (probability $P_\infty$, with the same approximation as
before). So it is successful with probability $p_rP_\infty=CP_\infty/r^{1+\sigma}$. Condition (A) says that at least one $\mathcal{A}$-link
of node $0$ is successful. This is the situation of Eq.~\eqref{eq:Pu} with $u=CP_\infty$, so,
\begin{equation}
    \mathrm{Prob}(\mathrm{A})=q(CP_\infty).
\end{equation}

\textbf{Condition (B).} The same as (A) but with the links of layer $\mathcal{B}$: at least one $\mathcal{B}$-link of node $0$ is successful, and the probability is $\mathrm{Prob}(\mathrm{B})=q(CP_\infty)$.

We apply both conditions simultaneously. The links of $\mathcal{B}$ are drawn independently of the links of $\mathcal{A}$, and in the tree approximation, the nodes reached through them are independent as well. The two probabilities multiply to give
\begin{equation}
    P_\infty=q(CP_\infty)^2=\left[1-P_+(CP_\infty)^2\right]^2.
    \label{eq:Pmul_tree}
\end{equation}
Compared with the single-layer case, the difference comes from the square of $q$.

As in the single-layer case, $P_\infty=0$ is the trivial solution of Eq.~\eqref{eq:Pmul_tree}. We find the other solutions using, again, $u=CP_\infty$ so that, Eq.~\eqref{eq:Pmul_tree} reads $P_\infty=q(u)^2$, and for every $u$ between $0$ and $1$, non-zero solutions exist at a given $C$ only if
$C=u/q(u)^2$ for some $u$.

Figure~\ref{fig:tree}(a) shows how the solutions are found graphically: a non-zero solution of Eq.~\eqref{eq:Pmul_tree} is an intersection of the curve $q(CP_\infty)^2$ with the diagonal. The right-hand side starts with zero slope, so for small $C$ the curve lies below the diagonal and only $P_\infty=0$ is a solution. A non-zero solution first appears when the curve touches the diagonal, at a finite value of $P_\infty$. Figure~\ref{fig:tree}(b) shows $C=u/q(u)^2$ as a function of $u$, and Fig.~\ref{fig:tree}(c) the resulting solutions $P_\infty(C)$. The circles in Fig.~\ref{fig:tree}(c) are results of simulations of the two-layer multiplex network itself, and not of a tree. They are described in Sec.~\ref{sec:sim}, and are shown here for comparison with the tree approximation. The function $u/q(u)^2$ has a minimum value for the following reason:

\begin{figure*}[t]
    \centering
    \includegraphics[width=0.92\textwidth]{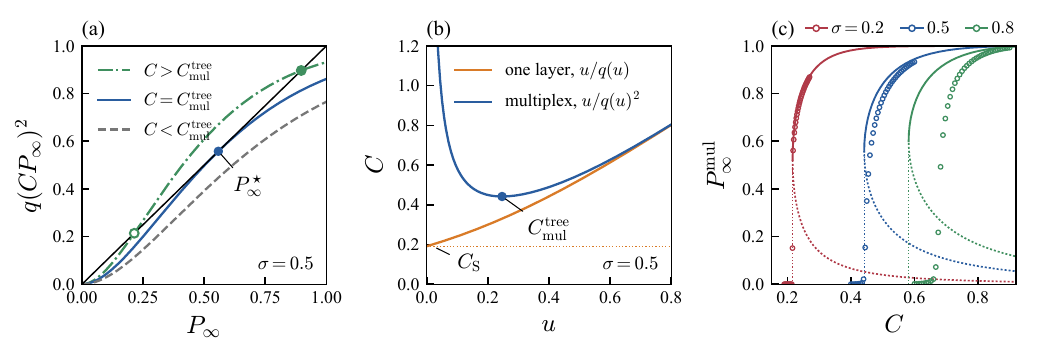}
    \caption{Tree approximation. (a) Graphical solution of Eq.~\eqref{eq:Pmul_tree} for $\sigma=0.5$. The right-hand side, $q(CP_\infty)^2$, starts with zero slope, so a non-zero solution (an intersection with the diagonal) exists only if $C\geq C^\mathrm{tree}_\mathrm{mul}$, where the curve touches the diagonal at $P_\infty^\star$. For $C>C^\mathrm{tree}_\mathrm{mul}$ the curve crosses the diagonal at two points: the stable solution (filled circle) and the unstable one (open circle). (b) The value of $C$ belonging to each $u=CP_\infty$, for one layer, Eq.~\eqref{eq:Cu}, and for the multiplex, $C=u/q(u)^2$. For one layer the minimum is at $u\to0$, where $C=C_\mathrm{S}$; for the multiplex the minimum, $C^\mathrm{tree}_\mathrm{mul}$, is at a finite $u^\star$. (c) Size of the mutual giant component as a function of $C$: tree approximation (solid lines: stable solution; dotted curves: unstable solution; vertical dotted lines: $C^\mathrm{tree}_\mathrm{mul}$) and simulation (circles). The simulation is of the two-layer multiplex network of Sec.~\ref{sec:model} itself, and not of a tree (Method~1 of Sec.~\ref{sec:sim}, $N=2^{16}$, 28 realizations).}
    \label{fig:tree}
\end{figure*}
\begin{itemize}
    \item For small $u$, Eq.~\eqref{eq:Pmul_tree} gives $C\approx u/\left(2\zeta(1+\sigma)u\right)^2=1/\left(4\zeta(1+\sigma)^2u\right)$, which is large.
    \item For large $u$, almost every node has a successful link, $q\to1$, and $C\approx u$, which grows.
\end{itemize}
In between, there is a lowest value, and no non-zero solutions exist below it:
\begin{equation}
    C^\mathrm{tree}_\mathrm{mul}(\sigma)=\min_{0<u<1}\frac{u}{q(u)^2}=\min_{0<u<1}\frac{u}{\left(1-P_+(u)^2\right)^2}.
    \label{eq:Ctree}
\end{equation}
If $u^\star$ is the position of the minimum value, the fraction of functional nodes right at the threshold is
\begin{equation}
    P_\infty^\star=q(u^\star)^2,
    \label{eq:Pstar}
\end{equation}
which is approximately $0.5$; it increases from $0.51$ for $\sigma\to0$ to $0.62$ at $\sigma=0.9$. For $C>C^\mathrm{tree}_\mathrm{mul}$ there are two non-zero solutions (two values of $u$ with the same $C$), which are marked by circles in Fig.~\ref{fig:tree}(a). The one with larger $P_\infty$ is the physical one: it is what one reaches by starting from $P_\infty=1$ and iterating $P\to q(u)^2$, which mimics the cascade of failures \cite{buldyrev2010,bashan2013}. The one with the smaller $P_\infty$ is unstable (dotted curves in Fig.~\ref{fig:tree}(c)).

The exact value of $C^\mathrm{tree}_\mathrm{mul}$ can be calculated by numerically minimizing Eq.~\eqref{eq:Ctree}, or by solving the following equation for $u^\star$. Setting the derivative of $\ln C=\ln u-2\ln q(u)$ to zero gives,
\begin{equation}
    u^\star q'(u^\star)=\frac{1}{2}q(u^\star),\quad q'(u)=2P_+(u)^2\sum_{r=1}^\infty \frac{1}{r^{1+\sigma}-u}.
    \label{eq:ustar}
\end{equation}

Unlike the single-layer case, $C^\mathrm{tree}_\mathrm{mul}$ is not a rigorous lower bound on $C_c^\mathrm{mul}$. In a single layer, the tree can only overestimate the number of nodes that are connected to node $0$, so its threshold can only be lower than the true one. In the multiplex, the tree approximation makes a second assumption: that the neighbors of node $0$ are functional independently of each other, and of node $0$. In the real multiplex, nearby nodes share many of their links, in both layers, and therefore they tend to fail together. These correlations are not controlled by the tree construction, and they can shift the threshold in either direction.

Although $C^\mathrm{tree}_\mathrm{mul}$ is only an approximation (and not a rigorous lower bound) for the multiplex $C_c^\mathrm{mul}$, for $\sigma\leq 0.5$ it is within $0.6\%$ of the actual value (see Table~\ref{tab:sim}). This is much better than the same approximation for the one-layer case, where $C_\mathrm{S}$ is within $1\%$ of $C_c^\mathrm{single}$ only for $\sigma=0.1$, and $25\%$ below it at $\sigma=0.5$. A plausible reading of this fact is that two errors nearly cancel: loops make each layer less connected than a tree, which raises the true threshold, while in the multiplex the failed nodes are clustered in space, which does less damage than randomly placed failures and lowers it.

\subsection{Numerical results}
\label{sec:sim}
We simulated the multiplex on rings of $N$ nodes with periodic boundary conditions, and located the threshold with two methods. Method~1 is based on the largest jump of the order parameter, and Method~2 on the number of iterations of the cascade of failures.

The distance between two nodes of a layer is their periodic distance on the ring, $1\le r\le N/2$. Testing each of the $N(N-1)/2$ pairs would take $O(N^2)$
operations. Instead, we use the rejection-free algorithm of Gori et al.~\cite{gori2017}: the $N$ pairs at a given
distance $r$ are linked independently with the same probability $p_r$, so the number of pairs skipped
between two consecutive links of length $r$ is geometrically distributed. Drawing these gaps generates the
links of each length in a time proportional to their number, and the whole network in $O(N)$ operations for
$\sigma>0$.

The cascade alternates
between the layers: every node outside the LCC of $\mathcal{A}$ fails together with its partner in
$\mathcal{B}$; then every node outside the LCC of $\mathcal{B}$ fails together with its partner in
$\mathcal{A}$; and so on, until no further node fails. The surviving nodes form the mutual giant component,
whose relative size is $P_\infty^\mathrm{mul}$. The LCC of a layer is found with the union--find algorithm~\cite{tarjan1975,newmanziff2000}, which goes over the links one by one and merges the clusters of the two nodes that each link connects. After every stage of the cascade, the LCC is recomputed from the links between the surviving nodes.

\textit{Method 1: size of the largest jump.}
In this method, the same realization of the network is followed for all the values of $C$. This is possible because the probability of every link is proportional to $C$. For each realization, we first generate the links of the network at the largest value of $C$ that we study, $C_{\max}$, and attach to every link an independent random number $m$, uniformly distributed between $0$ and $1$. The network at a smaller value, $C<C_{\max}$, is then obtained by keeping only the links with $m<C/C_{\max}$, and deleting all the others. Every link is kept with probability $C/C_{\max}$, so a link that exists with probability $p_r(C_{\max})$ at $C_{\max}$ exists with probability $p_r(C_{\max})\,C/C_{\max}=p_r(C)$ at $C$, exactly as in a network that is generated directly at $C$. When $C$ increases, links are added to the network but are never removed. Therefore, for each realization, $P_\infty^\mathrm{mul}(C)$ is effectively a non-decreasing function of $C$, which increases in steps. Small decreases occur occasionally, because the cascade follows the largest cluster of each layer, but they are small compared with the largest increase.

\begin{figure*}[t]
    \centering
    \includegraphics[width=\textwidth]{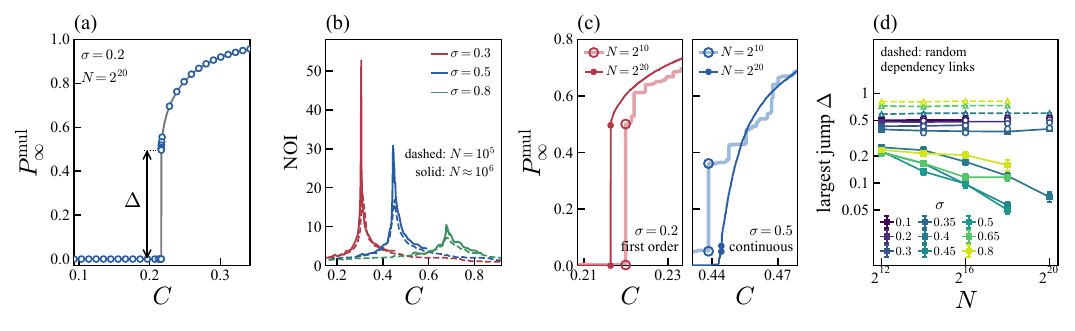}
    \caption{Simulations of the multiplex. (a) Method~1, shown for one realization with $\sigma=0.2$ and $N=2^{20}$. Line: $P_\infty^\mathrm{mul}(C)$ of the realization. Circles: the values of $C$ at which it was calculated. The height of the largest jump is $\Delta$, and its position is the estimate of $C_c^\mathrm{mul}$. (b) Method~2: number of cascade iterations (NOI) as a function of $C$, for $N=10^5$ (dashed lines, 30 realizations) and $N\approx10^6$ (solid lines, 10 realizations). The position of the maximum is the estimate of $C_c^\mathrm{mul}$. (c) $P_\infty^\mathrm{mul}(C)$ of single realizations with $N=2^{10}$ and $N=2^{20}$. The circles mark the two ends of the largest jump of each realization (open circles: $N=2^{10}$; filled circles: $N=2^{20}$). For $\sigma=0.2$ the mutual giant component appears in a single jump, of the same height for the two sizes (first-order transition). For $\sigma=0.5$ the largest jump decreases with $N$, and for the larger size the growth is gradual (continuous transition). For each $\sigma$ and $N$, the realization with the median largest jump among 21 realizations (10 for $N=2^{20}$) is shown. (d) Mean height $\Delta$ of the largest jump as a function of $N$ (Method~1). Open circles: the independent runs with $N=2^{14}$ to $2^{20}$ in which the two methods were compared (20 realizations, 10 at $N=2^{20}$), for $\sigma\le0.35$. For $\sigma\le0.35$, $\Delta$ is independent of $N$ (first-order
    transition); for $\sigma\ge0.4$ it decreases with $N$ (continuous transition). Dashed lines: the same layers with random dependency links, for $\sigma=0.4$, $0.65$ and $0.8$; here $\Delta$ is independent of $N$ (first-order transition).}
    \label{fig:sim}
\end{figure*}

We locate the largest step of this function as follows [Fig.~\ref{fig:sim}(a)]. We first calculate $P_\infty^\mathrm{mul}$ at 25 equally spaced values of $C$, which divide the range of $C$ into intervals, and find the interval over which $P_\infty^\mathrm{mul}$ increases the most. We divide this interval into two halves, by calculating $P_\infty^\mathrm{mul}$ at its middle point, and look again for the interval with the largest increase of $P_\infty^\mathrm{mul}$, among all the intervals. This is repeated until the interval with the largest increase is narrower than $10^{-4}C_{\max}$. The increase of $P_\infty^\mathrm{mul}$ over this last interval is the largest jump of the realization, $\Delta$, and the position of the jump, averaged over the realizations, is our estimate of $C_c^\mathrm{mul}$. The height of the jump distinguishes the two kinds of transition:
$\Delta$ tends to a non-zero constant as $N\to\infty$ at a first-order transition, and to zero at a
continuous one. We used $N=2^{12}$ to $2^{20}$ and 14 to 28 realizations per point.

\textit{Method 2: number of cascade iterations.}
Close to the threshold the cascade slows down, and the
number of iterations (NOI) needed to reach the final state has a sharp maximum. This was used to locate the threshold of first-order transitions in interdependent networks, where the NOI at the threshold grows as a power of the system size~\cite{parshani2011,zhou2014}. We record the NOI on a
grid of $C$, average it over the realizations, and take the position of the maximum as
$C_c^\mathrm{mul}$ [Fig.~\ref{fig:sim}(b)]. We used two system sizes: $N=10^5$, with 30 realizations and a grid spacing of $0.01$, and $N=2^{20}\approx10^6$, with 10 realizations and a grid spacing of $0.001$ around the maximum.

The cascade is an iterative process also when the transition is continuous~\cite{parshani2010}, and we use the NOI maximum for all the values of $\sigma$. To test it, we applied both methods to the same networks, with $N=2^{14}$ to $2^{20}$. At $N=2^{20}$, the thresholds of the two methods agree within $1\%$ for all $\sigma$ (Table~\ref{tab:sim}).

\textit{Finite-size effects at small $\sigma$.}
Because of the periodic boundary conditions, the distance between two nodes on the ring is at most $N/2$, so there are no links longer than $N/2$. The mean degree of a node is therefore
\begin{equation}
    \langle k\rangle_N=2C\sum_{r=1}^{N/2}r^{-(1+\sigma)},
    \label{eq:kN}
\end{equation}
which is smaller than the mean degree of the infinite chain, $\langle k\rangle=2C\zeta(1+\sigma)$ [Eq.~\eqref{eq:k}]. The links that are missing are those longer than $N/2$, and their share of the mean degree is
\begin{equation}
    \frac{\sum_{r>N/2}r^{-(1+\sigma)}}{\zeta(1+\sigma)}\simeq\frac{(N/2)^{-\sigma}}{\sigma\,\zeta(1+\sigma)},
    \label{eq:missing}
\end{equation}
where the sum was approximated by the integral $\int_{N/2}^\infty r^{-(1+\sigma)}\,\mathrm{d}r=(N/2)^{-\sigma}/\sigma$. This missing fraction is $32\%$ at $\sigma=0.1$ and $10\%$ at $\sigma=0.2$
for $N=10^5$, but only $1\%$ at $\sigma=0.4$.

The missing links shift the threshold that is measured on the ring. We denote the threshold of a ring of $N$ nodes by $C_c^\mathrm{mul}(N)$, and the threshold of the infinite chain by $C_c^\mathrm{mul}$, as before. At small $\sigma$ most of the links are long, and the layers are similar to random networks, for which the mutual giant component appears when the mean degree exceeds a certain value~\cite{buldyrev2010}. We therefore assume that the mean degree at the threshold is the same on the ring and in the infinite chain. On the ring, the mean degree at the threshold is given by Eq.~\eqref{eq:kN} with $C=C_c^\mathrm{mul}(N)$, and in the infinite chain it is given by Eq.~\eqref{eq:k} with $C=C_c^\mathrm{mul}$. Equating the two gives
\begin{equation}
    C_c^\mathrm{mul}(N)\simeq C_c^\mathrm{mul}\,
    \frac{\zeta(1+\sigma)}{\sum_{r=1}^{N/2}r^{-(1+\sigma)}}.
    \label{eq:truncation}
\end{equation}
The sum in the denominator is smaller than $\zeta(1+\sigma)$, so the threshold of the ring is higher than the threshold of the infinite chain. Equation~\eqref{eq:truncation} is an approximation, because the threshold of the network is not determined by its mean degree alone. For example, at $\sigma=0.1$ and $N=10^5$ it gives a threshold that is higher than $C_c^\mathrm{mul}$ by a factor of $1.47$.

Since the missing fraction decreases only as $N^{-\sigma}$, increasing $N$ is not a practical way to remove this shift at small $\sigma$. Instead, we add the missing links to the ring, in the following way. On a ring, the distance between two nodes is not unique. Besides the shortest distance $r$, one can also reach the second node by going around the ring in the other direction, or more than once, which gives the distances $N-r$, $N+r$, $2N-r$, and so on. In the infinite chain, these are the distances from the first node to the periodic images of the second node, namely the nodes that are shifted from it by a multiple of $N$. We take the probability of a link between two nodes of the ring as the sum of the probabilities of the links of all these lengths,
\begin{equation}
    p_r^{(N)}=C\sum_{n=-\infty}^{\infty}\frac{1}{|r+nN|^{1+\sigma}}.
    \label{eq:images}
\end{equation}
The term $n=0$ is the original probability $p_r$, and the other terms add the links that are longer than $N/2$. The mean degree is then $2C\zeta(1+\sigma)$, as in the infinite chain, up to a relative error $N^{-(1+\sigma)}$.

Equation~\eqref{eq:images} does not change the model in the limit $N\to\infty$. For every fixed $r$, $p_r^{(N)}\to p_r$, and each node has a link of length $\ell$ with probability $C/\ell^{1+\sigma}$ for all $\ell$, as in the infinite chain; only the end nodes of the links longer than $N/2$ are changed. The same sum over the periodic images is used in simulations of the Ising model with LR interactions, in order to reduce the finite-size corrections~\cite{horita2017}.

We used Eq.~\eqref{eq:images} in Method~1, and in Method~2 with $N\approx10^6$. The runs of Method~2 with $N=10^5$ do not include the periodic images, and their thresholds are therefore shifted at small $\sigma$: they are $0.17$ and $0.24$ at $\sigma=0.1$ and $0.2$, compared with $0.115$ and $0.216$ with the periodic images (Table~\ref{tab:sim}). To verify that this difference is due to the missing links, and not to the method, we also ran Method~1 with $N=10^5$ and without the periodic images. The thresholds are $0.169$ and $0.241$, in agreement with those of Method~2. The shift agrees with Eq.~\eqref{eq:truncation}: when the thresholds found with $N=10^5$ are divided by the factor of Eq.~\eqref{eq:truncation}, they agree with the others within their grid spacing of $0.01$ for all $\sigma$.

\begin{figure}[t]
    \centering
    \includegraphics[width=\columnwidth]{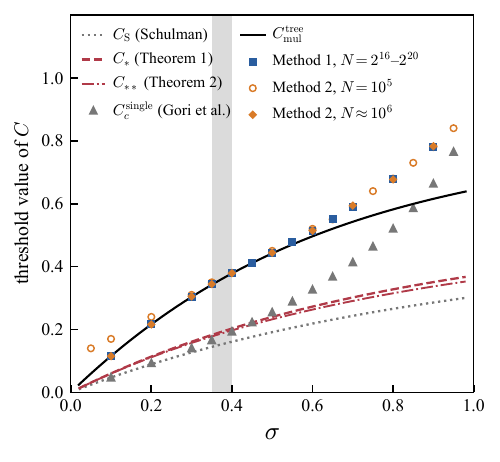}
    \caption{Critical values of $C$ as functions of $\sigma$. Lines: the bounds $C_\mathrm{S}$, $C_*$, $C_{**}$ and the
    tree approximation $C^\mathrm{tree}_\mathrm{mul}$. Triangles: single-layer threshold~\cite{gori2017}. Symbols:
    simulated multiplex threshold from Method~1 (squares; largest size simulated, $N=2^{16}$ to $2^{20}$), from Method~2 with $N=10^5$ (circles; the deviation at small $\sigma$ is the finite-size
    effect of Eq.~(\ref{eq:truncation})), and from Method~2 with $N\approx10^6$ (small diamonds). Gray band: $0.35<\sigma<0.4$, where the simulations indicate that the transition changes from first order (smaller $\sigma$) to continuous (larger $\sigma$).}
    \label{fig:thresholds}
\end{figure}

The runs with $N=10^5$ also included the two end points, $\sigma=0$ and $\sigma=1$, where the finite-size effect is extreme (these two points are not included in Table~\ref{tab:sim} and Fig.~\ref{fig:thresholds}). At $\sigma=0$ the sum $\sum_rr^{-1}$ diverges. The mean degree of the infinite chain is then infinite for every $C>0$, so that $C_c^\mathrm{mul}=0$, while on a finite ring the mean degree, Eq.~\eqref{eq:kN}, is finite and grows only as $\ln N$. For $N=10^5$ the NOI maximum is at $C\approx0.11$, where Eq.~\eqref{eq:kN} gives $\langle k\rangle_N\approx2.5$. This is close to $2.4554$, the mean degree at the threshold of two interdependent random networks~\cite{buldyrev2010}, as expected from the assumption that led to Eq.~\eqref{eq:truncation}. At $\sigma=1$ a single layer does not percolate for any $C<1$~\cite{newman1986,aizenman1986}, so
$C_c^\mathrm{mul}=1$, and the weak NOI maximum found between $C=0.85$ and $0.9$ for $N=10^5$ is again a finite-size value.

\begin{table*}[t]
    \caption{Critical values of $C$ as a function of $\sigma$. $C_\mathrm{S}$: Schulman's bound for a single layer, Eq.~\eqref{eq:CS}; $C_{**}$ and $C_*$: the
    rigorous lower bounds of Theorems~2 and~1; $C_c^\mathrm{single}$: single-layer threshold from
    Ref.~\cite{gori2017}; $C^\mathrm{tree}_\mathrm{mul}$: tree approximation, Eq.~\eqref{eq:Ctree}. Simulations of the multiplex:
    Method~1 (largest jump, with periodic images, largest size simulated, $N=2^{16}$ to $2^{20}$), and Method~2 (NOI maximum), for $N=10^5$ (grid spacing $0.01$, without periodic images) and for $N\approx10^6$ (grid spacing $0.001$, with periodic images). $\Delta$ is the height of the largest
    jump; ``$\to0$'' means that it decreases with $N$ [Fig.~\ref{fig:sim}(d)]. Last two columns: threshold and largest jump of
    Method~1 when the dependency links connect randomly chosen pairs of nodes; here $\Delta$ is independent of $N$ for all $\sigma$.}
    \label{tab:sim}
    \begin{ruledtabular}
    \begin{tabular}{cccccccccccc}
    & \multicolumn{5}{c}{analytical} & \multicolumn{4}{c}{multiplex, simulation} & \multicolumn{2}{c}{random links}\\
    $\sigma$ & $C_\mathrm{S}$ & $C_{**}$ & $C_*$ & $C_c^\mathrm{single}$ & $C^\mathrm{tree}_\mathrm{mul}$
    & Method 1 & $\Delta$ & Method 2 & Method 2 & Method 1 & $\Delta$\\
    & (Schulman~\cite{schulman1983}) & & & (Gori \textit{et al.}~\cite{gori2017}) & & & & $N=10^5$ & $N\approx10^6$ & & \\
    \hline
    0.1  & 0.0472 & 0.0596 & 0.0603 & 0.0477 & 0.1155 & 0.1153 & 0.51 & 0.17 & 0.115 & 0.1160 & 0.52\\
    0.2  & 0.0894 & 0.1116 & 0.1138 & 0.0932 & 0.2164 & 0.2162 & 0.49 & 0.24 & 0.216 & 0.2207 & 0.54\\
    0.3  & 0.1272 & 0.1571 & 0.1611 & 0.1405 & 0.3035 & 0.3039 & 0.44 & 0.31 & 0.305 & 0.3167 & 0.57\\
    0.35 & 0.1445 & 0.1777 & 0.1826 & 0.1661 & 0.3423 & 0.3442 & 0.40 & 0.35 & 0.344 & 0.3630 & 0.58\\
    0.4  & 0.1610 & 0.1971 & 0.2029 & 0.1935 & 0.3781 & 0.3793 & $\to0$ & 0.38 & 0.379 & 0.4077 & 0.60\\
    0.5  & 0.1914 & 0.2322 & 0.2399 & 0.2548 & 0.4419 & 0.4433 & $\to0$ & 0.45 & 0.444 & 0.4953 & 0.65\\
    0.6  & 0.2187 & 0.2633 & 0.2727 & 0.3271 & 0.4963 & 0.514  & $\to0$ & 0.52 & 0.514 & 0.5801 & 0.70\\
    0.7  & 0.2434 & 0.2908 & 0.3018 & 0.4138 & 0.5425 & 0.590  & $\to0$ & 0.59 & 0.594 & 0.6650 & 0.76\\
    0.8  & 0.2656 & 0.3152 & 0.3277 & 0.5210 & 0.5819 & 0.677  & $\to0$ & 0.68 & 0.677 & 0.7534 & 0.82\\
    0.9  & 0.2858 & 0.3369 & 0.3507 & 0.6641 & 0.6154 & 0.780  & $\to0$ & 0.78 & 0.783 & 0.8461 & 0.89\\
    \end{tabular}
    \end{ruledtabular}
\end{table*}

\textit{The critical value.}
Table~\ref{tab:sim} and Fig.~\ref{fig:thresholds} summarize the thresholds. The simulated
$C_c^\mathrm{mul}$ lies above all the lower bounds and above the single-layer threshold at every $\sigma$. The
coupling raises the threshold by a factor $C_c^\mathrm{mul}/C_c^\mathrm{single}=2.4$ at $\sigma=0.1$,
decreasing to $1.17$ at $\sigma=0.9$; for $\sigma\to0$ the factor approaches $2.4554$, the ratio of the
thresholds of two interdependent and one single Erd\H{o}s--R\'enyi network~\cite{buldyrev2010}. The rigorous bounds $C_*$ and
$C_{**}$ are about half of the simulated values, and for $\sigma\gtrsim0.45$ they fall below
$C_c^\mathrm{single}$, so that there the trivial bound $C_c^\mathrm{mul}\ge C_c^\mathrm{single}$ is
stronger.

The tree approximation $C^\mathrm{tree}_\mathrm{mul}$ is within $0.6\%$ of the simulated value
for $\sigma\le0.5$. For larger $\sigma$ it is too low, by $3.5\%$ at $\sigma=0.6$ and by $21\%$ at $\sigma=0.9$, where short links
dominate and the network is far from a tree. We note that $C^\mathrm{tree}_\mathrm{mul}$ is not always lower than the simulated value: at $\sigma=0.1$ it is higher by $0.1\%$ ($0.11549$, compared with $0.11535\pm0.00003$ from Method~1 with $N=2^{20}$), and at $\sigma=0.2$ the two values are equal within the statistical error. The simulations therefore do not suggest that $C^\mathrm{tree}_\mathrm{mul}$ is a lower bound for $C_c^\mathrm{mul}$, in agreement with the discussion of Sec.~\ref{sec:tree}.

\textit{Order of the transition.}
Figure~\ref{fig:tree}(c) shows that the averaged order parameter, $P_\infty^\mathrm{mul}$, rises abruptly for all $\sigma$, but an abrupt rise
at a single system size does not decide the order of the transition. To determine the order, we examine the steps of $P_\infty^\mathrm{mul}(C)$ in single realizations, and how they depend on $N$. Figure~\ref{fig:sim}(c) shows $P_\infty^\mathrm{mul}(C)$ of single realizations for two values of $\sigma$, each for two system sizes, $N=2^{10}$ and $N=2^{20}$. For $\sigma=0.2$, the mutual giant component appears in a single jump, from zero to about half of the nodes, and the height of the jump is the same for the two sizes ($0.50$), although they differ by a factor of $1000$. For $\sigma=0.5$, in contrast, the largest jump decreases with $N$, from $0.31$ for $N=2^{10}$ to $0.02$ for $N=2^{20}$. For the larger size the mutual giant component grows gradually, in many small steps. A gradual growth of this kind is characteristic of a continuous transition, as in interdependent lattices with short dependency links~\cite{li2012}. Figure~\ref{fig:sim}(d) shows how the
mean largest jump, $\Delta$, depends on $N$. For $\sigma\le0.35$ it is independent of $N$, between $0.40$ and $0.51$:
the transition is first order. The runs with $N=2^{14}$ to $2^{20}$ in which the two methods were compared confirm this [open circles in Fig.~\ref{fig:sim}(d)]: at $N=2^{20}$, $\Delta=0.53$, $0.51$, $0.47$ and $0.41$ for $\sigma=0.1$, $0.2$, $0.3$ and $0.35$, the same as, or slightly larger than, at $N=2^{14}$ ($0.52$, $0.50$, $0.43$ and $0.36$). At small $\sigma$ the jump is close to the tree value
$P_\infty^\star\approx0.5$, and it decreases as $\sigma$ grows. For $\sigma\ge0.4$, $\Delta$ decreases with
$N$, by a factor of $3.6$ to $4.5$ over the sizes studied for $\sigma=0.4$--$0.5$, with no sign of
saturation: the transition is continuous. The fitted decay exponents, $-\mathrm{d}\ln\Delta/\mathrm{d}\ln N$ between $0.06$ and
$0.39$ with large statistical errors, scatter around the value $(1-\sigma)/2$ expected for a continuous
transition whose critical clusters have the fractal dimension $(1+\sigma)/2$ of single-layer long-range
percolation~\cite{hutchcroft2025II}. The order of the transition therefore changes at
$0.35<\sigma_\times<0.4$. As $\sigma\to1$ the decay becomes very slow ($N^{-0.05}$ at $\sigma=0.9$), so that
there the transition looks first order at any accessible size; this mirrors the discontinuity of the
single-layer transition at $\sigma=1$~\cite{aizenman1986,duminil2024}.

We now return to the NOI. It has a maximum both for $\sigma\leq0.35$, where the transition is first order, and for $\sigma\geq0.4$, where it is continuous [Fig.~\ref{fig:sim}(b)], and in both cases its position agrees with Method~1 (Table~\ref{tab:sim}). The NOI maximum therefore detects the critical value for both kinds of transition. The height of the maximum, however, cannot be measured on a grid of $C$. At a first-order transition the NOI rises sharply as $C$ approaches the threshold of the realization, in a range of $C$ that becomes narrower as $N$ grows~\cite{parshani2011,zhou2014}, so the value found on a grid depends on the distance of the nearest grid point from the threshold. We therefore measured the NOI at the threshold of each realization. For this purpose, the halving of Method~1 was continued until the networks at the two ends of the interval of the largest jump differ by a single link, and we recorded the NOI of the last cascade before the jump. In randomly interdependent networks, the NOI at the threshold of each realization grows as $N^{1/3}$~\cite{zhou2014}. We find the same law for $\sigma\leq0.3$: for $N=2^{12}$ to $2^{20}$ the mean NOI grows as $N^x$, with $x=0.34\pm0.01$, $0.33\pm0.01$ and $0.35\pm0.01$ for $\sigma=0.1$, $0.2$ and $0.3$. For $\sigma\geq0.5$, where the transition is continuous, the NOI grows much more slowly, with $x=0.20$, $0.14$ and $0.13$ for $\sigma=0.5$, $0.6$ and $0.8$ (for $\sigma=0.8$ the largest size is $N=2^{18}$). It grows fastest near $\sigma_\times$, with $x\approx0.4$ at $\sigma=0.35$ and $0.4$.

\textit{Random dependency links.}
As a control, we repeated the simulations of Method~1 with the dependency links connecting node $i$ of
$\mathcal{A}$ to a randomly chosen node of $\mathcal{B}$, one to one. This is the setting of
Refs.~\cite{buldyrev2010,bashan2013}, in which the failures passed from one layer to the other are randomly placed. The results differ from those of the multiplex in two ways (last two columns of Table~\ref{tab:sim}).

First, the threshold with random dependency links is higher than the threshold of the multiplex, by $0.6\%$ at $\sigma=0.1$ and by $9$--$13\%$ for $\sigma\ge0.5$. This supports the interpretation of Sec.~\ref{sec:tree}: in the multiplex the failed
nodes are clustered in space, which does less damage than randomly placed failures. It also shows how large the two errors of the tree approximation are. With random dependency links the failures are randomly placed, as the tree approximation assumes, and only the loops are neglected. The threshold is then above $C^\mathrm{tree}_\mathrm{mul}$ for all $\sigma$, by $0.4\%$ at $\sigma=0.1$ and by $12\%$ at $\sigma=0.5$. For $\sigma\le0.5$, the clustering of the failures in the multiplex lowers the threshold by almost the same amount, so that the two errors nearly cancel.

Second, with random dependency links the transition is first order for all $\sigma$. The largest jump $\Delta$ is independent of $N$ [dashed lines in Fig.~\ref{fig:sim}(d)]: between $N=2^{14}$ and the largest size simulated, it changes by less than $3\%$ for every $\sigma$. It increases from $0.52$ at $\sigma=0.1$ to $0.89$ at $\sigma=0.9$ (Table~\ref{tab:sim}). The multiplex thus interpolates between the two known cases of
interdependent lattices: first order when the dependency links are random~\cite{bashan2013}, continuous when they connect
nodes at the same position~\cite{son2011,li2012}.

\section{Discussion}
\label{sec:discussion}
We studied how long-range links affect the stability of interdependent spatial networks, in the simplest setting: a multiplex of two one-dimensional networks, in which the probability of a link decays as $C/r^{1+\sigma}$. The exponent $\sigma$ controls the range of the links, and we found that it determines both the critical value and the order of the transition. The dependency between the layers raises the critical value $C_c^\mathrm{mul}$ above the single-layer value $C_c^\mathrm{single}$, by a factor that decreases from $2.4$ at $\sigma=0.1$ to $1.17$ at $\sigma=0.9$. As $\sigma\to0$, almost all links are long, and the factor approaches the value $2.4554$ of two interdependent random networks~\cite{buldyrev2010}.

We presented two analytical estimates of $C_c^\mathrm{mul}$. The first, $C_*$, is a rigorous lower bound, based on counting the paths in one layer whose nodes also have links in the other layer. It is not sharp, and it is about half of the simulated value. The second is the tree approximation, $C^\mathrm{tree}_\mathrm{mul}$, which agrees with the simulations within $0.6\%$ for $\sigma\leq0.5$, but is not a rigorous bound. We do not have a full explanation for this agreement. The simulations with random dependency links suggest that two errors of the tree approximation nearly cancel: the loops, which reduce the connectivity of each layer, and the spatial clustering of the failures in the multiplex, which reduces their damage.

We note that the tree approximation predicts a first-order transition for all $\sigma$, with a jump $P_\infty^\star$ between $0.5$ and $0.6$ [Eq.~\eqref{eq:Pstar}]. It therefore cannot describe the continuous transition that we find for $\sigma\geq0.4$. Its estimate of the critical value, however, remains within $0.6\%$ of the simulated value up to $\sigma=0.5$, where the transition is already continuous, and it becomes inaccurate only for larger $\sigma$. The jump that it predicts, in contrast, deviates from the simulated one already for $\sigma<\sigma_\times$ ($0.54$ compared with $0.40$ at $\sigma=0.35$). Whether the failure of the tree approximation at large $\sigma$ and the change in the order of the transition have the same origin is an open question.

Perhaps the most interesting result is that the order of the transition depends on $\sigma$. For $\sigma\leq0.35$ the transition is first order, as for interdependent random networks~\cite{buldyrev2010}, while for $\sigma\geq0.4$ it is continuous, as for a multiplex of lattices~\cite{son2011,li2012}. The duration of the cascade supports this picture. A sharp increase of the NOI near the threshold is a characteristic of the first-order transition of interdependent networks~\cite{parshani2011,bashan2013}, and for $\sigma\leq0.3$ we find that the NOI at the threshold grows as $N^{1/3}$, as in randomly interdependent networks~\cite{zhou2014}, while for $\sigma\geq0.5$ it grows much more slowly. When the dependency links connect random pairs of nodes, the transition is first order for all $\sigma$, as for randomly interdependent lattices~\cite{bashan2013}. Here, $\sigma$ plays a role similar to the length of the dependency links in Ref.~\cite{li2012}, and to the characteristic length of the links in the spatial multiplex of Ref.~\cite{danziger2016}, where longer links also change the transition from continuous to discontinuous. We note that the value $0.35<\sigma_\times<0.4$ at which the order of the transition changes is close to $\sigma=1/3$, where the critical behavior of a single layer changes from mean field ($\sigma<1/3$) to non-mean field~\cite{hutchcroft2024,hutchcroft2025II} (see also Ref.~\cite{son2024poster}). Whether this is a coincidence is an open question.

We expect that the methods presented here can be extended to partial dependency, where only a fraction of the nodes have dependency links~\cite{parshani2010}, and to LR networks in higher dimensions~\cite{amit2023}. We plan to study these, as well as the critical exponents of the continuous transition for $\sigma>\sigma_\times$, in future work.

\begin{acknowledgments}
This work was supported by the United States--Israel Binational Science Foundation (BSF), Jerusalem, Israel (Grant No.~2020255). G.~A. thanks Idan Rooze for fruitful discussions. AI tools were used to assist in the analytical proofs, in some of the simulations, and in the editing of the manuscript. All the results were checked and verified by the authors.
\end{acknowledgments}

\appendix

\section{Proof of the lower bounds}
\label{app:bounds}
Here we prove Theorems~\ref{thm:bound} and~\ref{thm:closed}, using the quantities of Eqs.~\eqref{eq:k}--\eqref{eq:eta}.

\textit{Self-sustaining sets.} For a set of nodes $S$, let $\mathcal{A}[S]$ and $\mathcal{B}[S]$ be the subgraphs of the two layers that contain only the nodes of $S$ and the links between them. We call $S$ \textit{self-sustaining} if every node of $S$ belongs to an infinite connected component of $\mathcal{A}[S]$ and to an infinite connected component of $\mathcal{B}[S]$. A union of self-sustaining sets is self-sustaining, so there is a largest one, $M$, and every infinite cluster of functional nodes is contained in $M$. By translation invariance, a MGC exists only if $\mathrm{Prob}(0\in M)>0$.

\textit{Step 1: reduction to one path.} Suppose that $0\in M$. The component of node $0$ in $\mathcal{A}[M]$ is infinite, and every node has a finite number of links, so this component contains an infinite self-avoiding path $0=x_0,x_1,x_2,\dots$ (K\"onig's lemma). Every $x_k$ belongs to $M$, and therefore has at least one link in layer $\mathcal{B}$. Let $D_x=\{\deg_\mathcal{B}(x)\geq1\}$ be the event that node $x$ has at least one link in layer $\mathcal{B}$, where $\deg_\mathcal{B}(x)$ is the number of links of $x$ in $\mathcal{B}$, and $\Phi(X)=\mathrm{Prob}\left(\bigcap_{x\in X}D_x\right)$ for a finite set of nodes $X$. Let $\Gamma_n$ be the set of self-avoiding paths $\gamma$ of $n$ steps that start at node $0$, and $V(\gamma)$ the set of $n+1$ nodes of $\gamma$. The two layers are independent, so for every $n$,
\begin{equation}
    \mathrm{Prob}(0\in M)\leq\sum_{\gamma\in\Gamma_n}\mathrm{Prob}(\gamma\subseteq\mathcal{A})\,\Phi(V(\gamma)).
    \label{eq:A1}
\end{equation}

\textit{Step 2: the key lemma.} For every finite non-empty set $X$,
\begin{equation}
    \Phi(X)\leq q\,\eta^{|X|-1}.
    \label{eq:A2}
\end{equation}
To prove it, let $z=\min X$ and $Y=X\setminus\{z\}$, and assume $Y\neq\emptyset$ (for $Y=\emptyset$, $\Phi(X)=q$). For $y\in Y$, let $D_y^{-z}$ be the event that $y$ has a $\mathcal{B}$-link to a node other than $z$, $q_y^{-z}=\mathrm{Prob}(D_y^{-z})$, $\Phi_{-z}(W)=\mathrm{Prob}\left(\bigcap_{y\in W}D_y^{-z}\right)$, and $p_{zy}=p_{|z-y|}$.

(a) If node $z$ has no $\mathcal{B}$-link, which happens with probability $1-q$, then $D_y=D_y^{-z}$ for every $y\in Y$, and the events $D_y^{-z}$ do not depend on the links of $z$. Hence
\begin{equation}
    \Phi(X)=\Phi(Y)-(1-q)\,\Phi_{-z}(Y).
\end{equation}

(b) Let $S\subseteq Y$ be the set of nodes linked to $z$ in $\mathcal{B}$. Given $S$, $D_y$ holds for $y\in S$, and equals $D_y^{-z}$ for $y\notin S$, so
\begin{equation}
    \Phi(Y)=\sum_{S\subseteq Y}\prod_{y\in S}p_{zy}\prod_{y\in Y\setminus S}(1-p_{zy})\,\Phi_{-z}(Y\setminus S).
\end{equation}
The events $D_y^{-z}$ are increasing, so by the Harris--FKG inequality $\Phi_{-z}(Y)\geq\Phi_{-z}(Y\setminus S)\prod_{y\in S}q_y^{-z}$, and therefore
\begin{equation}
\begin{split}
    \Phi(Y)&\leq\Phi_{-z}(Y)\prod_{y\in Y}\left(1-p_{zy}+\frac{p_{zy}}{q_y^{-z}}\right)\\
    &=\Phi_{-z}(Y)\prod_{y\in Y}\frac{1-p_{zy}}{1-p_{zy}/q}.
\end{split}
\end{equation}
The last equality uses $1-q_y^{-z}=(1-q)/(1-p_{zy})$, which follows from translation invariance.

(c) Combining (a) and (b),
\begin{equation}
    \Phi(X)\leq\Phi(Y)\left[1-(1-q)\prod_{y\in Y}\frac{1-p_{zy}/q}{1-p_{zy}}\right].
\end{equation}
Since $z=\min X$, the distances $y-z$ are distinct positive integers, and every factor of the product is smaller than $1$. The product is therefore larger than the product over all distances $r\geq1$, which equals $\prod_r(1-p_r/q)/P_+$. Using $1-q=P_+^2$, the square brackets are at most $1-P_+\prod_r(1-p_r/q)=\eta$. So $\Phi(X)\leq\eta\,\Phi(X\setminus\{\min X\})$, and iterating down to a single node proves Eq.~\eqref{eq:A2}.

\textit{Step 3: conclusion.} The expected number of self-avoiding paths of $n$ steps from node $0$ in layer $\mathcal{A}$ is at most $\langle k\rangle^n$. With Eqs.~\eqref{eq:A1} and~\eqref{eq:A2},
\begin{equation}
    \mathrm{Prob}(0\in M)\leq q\left(\langle k\rangle\eta\right)^n\xrightarrow[n\to\infty]{}0\quad\text{if}\quad\langle k\rangle\eta<1.
\end{equation}
This is the argument of Schulman~\cite{schulman1983}, in which every step of the path costs a factor $\langle k\rangle$, with an additional factor $\eta$ for the links of its new node in the other layer.

\textit{Step 4: monotonicity in $C$.} Each factor $(1-Cr^{-1-\sigma})^2$ is non-negative, decreasing and convex in $C$, so $1-q$ is convex and $q$ is concave, with $q(0)=0$. Hence $C/q$, and therefore $p_r/q$, are non-decreasing in $C$, every factor $(1-p_r)(1-p_r/q)$ is strictly decreasing, and $\langle k\rangle\eta$ is strictly increasing. As $C\to0$, $\langle k\rangle\eta\to0$, and as $C\to1$, $p_1\to1$, $\eta\to1$ and $\langle k\rangle\eta\to2\zeta(1+\sigma)>2$. This proves Theorem~\ref{thm:bound}.

\textit{Theorem~\ref{thm:closed}.} The function $\ln(1-x)$ is concave, so on $[0,c]$ it lies above its chord, $\ln(1-x)\geq(x/c)\ln(1-c)$. With $c=C$ this gives $P_+\geq(1-C)^z$, and with $c=C/q$ it gives $\prod_r(1-p_r/q)\geq(1-C/q)^z$, where $z=\zeta(1+\sigma)$ and we used $\sum_rp_r=Cz$. Finally, $q\geq1-e^{-2Cz}$, since $1-p\leq e^{-p}$, and $1-C/q$ increases with $q$. Therefore
\begin{equation}
    \eta\leq1-\left[(1-C)\left(1-\frac{C}{1-e^{-2Cz}}\right)\right]^z,
\end{equation}
which proves Theorem~\ref{thm:closed}.

\textit{The cascade of failures.} In an infinite system, the set of nodes that survive the cascade of failures need not be self-sustaining. However, Step~1 only uses the fact that node $0$ has an infinite path in layer $\mathcal{A}$ whose nodes have at least one link in layer $\mathcal{B}$. This already holds after the first three stages of the cascade. After the second stage, every surviving node has at least one link in $\mathcal{B}$, but its path in $\mathcal{A}$ may have been broken by the nodes that failed in that stage. A node that also survives the third stage has an infinite path in $\mathcal{A}$ that consists of survivors of the second stage. Therefore, under the condition of Theorem~\ref{thm:bound} the cascade also leaves no functional nodes.

\textit{Random dependency links.} If node $x$ of $\mathcal{A}$ depends on node $\pi(x)$ of $\mathcal{B}$, for any one-to-one map $\pi$, Step~1 requires that the nodes $\pi(x_k)$ have links in $\mathcal{B}$. These nodes are distinct, and Eq.~\eqref{eq:A2} holds for every finite set, so Theorems~\ref{thm:bound} and~\ref{thm:closed} hold for any fixed one-to-one map $\pi$ between the nodes of the two layers.

\bibliography{references}

\end{document}